\documentclass[%
 aip,
rsi,%
 amsmath,amssymb,
 reprint,%
]{revtex4-1}

\usepackage{xcolor}
\usepackage{graphicx}% Include figure files
\usepackage{dcolumn}% Align table columns on decimal point
\usepackage{bm}% bold math
\usepackage{nicefrac}
\begin{document}

\preprint{AIP/123-QED}

\title[Rodr\'{i}guez et al.]{Estimating coil features from an equilibrium}% Force line breaks with \\

\author{E. Rodr\'{i}guez}
\affiliation{ Max Planck Institute for Plasma Physics, 17491 Greifswald, Germany}
 \altaffiliation[Email: ]{eduardo.rodriguez@ipp.mpg.de}

\author{W. Sengupta}%
\affiliation{%
Department of Astrophysical Sciences, Princeton University, Princeton, NJ 08543, USA 
%\\This line break forced% with \\
} 

\date{\today}% It is always \today, today,

\begin{abstract}
We present an explicit theoretical framework for constructing artificial modular coils { for vacuum stellarator fields} based solely on equilibrium properties, achieved through the formulation of a current potential defined on flux surfaces. Contours of constant Boozer toroidal angle can be directly interpreted as proxy coils, and so we demonstrate that key measures of coil complexity --particularly coil non-planarity-- are strongly governed by local magnetic field properties. { This approach} shows promise as predictor for more realistic coil configurations, providing both a pathway towards deeper understanding of equilibrium–coil relationships and a potential practical proxy for coil design.
\end{abstract}

\maketitle

% \section{\label{sec:intro} Introduction}
The stellarator \cite{spitzer1958,boozer1998,Helander2014} is an attractive magnetic confinement concept for achieving controlled thermonuclear fusion. Its flexibility and generality make it unique in its ability to confine plasmas while avoiding common shortcomings of other confinement designs \cite{schuller1995,boozer1998}. However, achieving good confinement in stellarators requires careful tailoring of the magnetic field geometry \cite{mynick2006}. In practice, reproducing three-dimensional (3D) stellarator magnetic fields requires coils, or other current distributions, of considerable geometric complexity. This imposes a significant burden on stellarator design and necessitates careful consideration within the design process itself. A natural question therefore arises: which stellarator fields are particularly amenable to simple coils?
\par
A central difficulty in addressing such a question is the fundamental lack of uniqueness of the inverse coil problem: the same magnetic field within a volume can, in principle, be reproduced to arbitrary accuracy by many external current distributions.\cite{lima2006magnetic, landreman2017improved} Thus, in practice, the problem is commonly approached as an optimisation task in which the field error is minimised. Despite substantial progress, particularly in optimisation strategies and regularisation techniques, \cite{merkel1987solution,landreman2016,drevlak2019optimisation,gates2017recent,zhu2018new} an important question remains: to what extent is a given current distribution solution intrinsic to the target magnetic field, and to what extent is it a consequence of the optimisation procedure and the imposed constraints? 
\par
Recent insight into this question has been gained through the analysis of intrinsic magnetic field length scales, such as $L_{\nabla\mathbf{B}}$.\cite{landreman2021a,kappel2024magnetic} However, such measures remain one step removed from the coil design problem, as they do not directly describe properties of the current distributions. In this Letter, we exploit a theoretical construction by which a coil set can be defined directly and uniquely for a given stellarator { vacuum} equilibrium. The construction is based on a surface current defined on a magnetic flux surface, thereby removing the inherent non-uniqueness of the coil problem. { The resulting infinite coil-set interpretation of such current distribution identifies coil shapes with constant Boozer toroidal angle contours.} Although the resulting coil set is artificial, it directly encodes intrinsic properties of the magnetic field that can be leveraged to assess the behaviour of more realistic coil sets. We therefore propose it as a practical tool for coil assessment and as a framework for gaining deeper theoretical insight into stellarator coil design.

\textit{Constructing the coil set.}
We begin from the standard Merkel formulation\cite{merkel1987solution} of the coil problem. Let $\mathbf{B}$ be a vacuum magnetic field in a toroidal volume $\Omega$ (we leave the finite plasma $\beta$ problem to future work), whose boundary $\partial\Omega$ a flux surface, $\mathbf{B}\cdot\hat{\mathbf{n}}=0$, with $\hat{\mathbf{n}}$ the surface normal. Introducing a toroidal winding surface $S$ enclosing $\Omega$, the Merkel problem consists of finding a scalar potential $\Phi(\theta,\phi)$ as a function of the poloidal and toroidal angles, respectively, such that the surface current

\begin{equation}
    \mathbf{K}=\frac{1}{\mu_0}\hat{\mathbf{n}}\times\nabla\Phi,
\end{equation}
reproduces $\mathbf{B}$ in $\Omega$. In practice, this inverse problem is solved numerically via least-squares minimisation (e.g. \texttt{NESCOIL}\cite{merkel1987solution}, \texttt{REGCOIL}\cite{landreman2017improved}, \texttt{QUADCOIL}\cite{fu2025global}). Different solutions obtained in this way depend on the imposed regularisation \cite{fu2025global} and the choice of $S$.

The lack of uniqueness is due to the finite volume between $S$ and $\partial\Omega$, but it can be eliminated by considering instead the limiting case, $S \to \partial\Omega$. In this limit, the winding surface coincides with a flux surface, and by direct application of the virtual casing principle\cite{drevlak2005pies,hanson2015virtual}, we have a closed form for the surface current

\begin{equation}
    \mathbf{K}=-\frac{1}{\mu_0}\hat{\mathbf{n}}\times\mathbf{B},
    \label{eqn:K_anal_B}
\end{equation}
evaluated on $\partial\Omega$. The surface current is simply orthogonal to the magnetic field everywhere and therefore introduces a discontinuity in its tangential component. It can be seen as a special case of [\onlinecite{merkel1986integral}].

That the above current reproduces $\mathbf{B}$ follows directly from the vanishing of the exterior field. Applying Biot–Savart and writing $\mathbf{B}=-\nabla\Phi_0$ in $\Omega$, one obtains
\begin{align}
\mathbf{B}_{\bar{\Omega}}
&= \frac{\mu_0}{4\pi}\int_S \mathbf{K}\times\nabla'\left(\frac{1}{\rho}\right)\mathrm{d}S' \\
&= \frac{1}{4\pi}\int_S \hat{\mathbf{n}}\cdot\mathbf{B} \, \nabla'\left(\frac{1}{\rho}\right) \mathrm{d}S',
\end{align}
% where we used $\nabla^2(1/\rho)=-4\pi\delta(\mathbf{r}-\mathbf{r}')$.
which vanishes identically since $\mathbf{B}\cdot\hat{\mathbf{n}}=0$ on a flux surface. {Here $\rho=|\mathbf{r}-\mathbf{r}'|$ and the primed quantities denote integration or derivatives with respect to $\mathbf{r}'$.} The tangential discontinuity thus enforces $\mathbf{B}$ throughout $\Omega$. It is therefore imperative that $S$ be a flux surface, since the problem in which $S$ and $\partial\Omega$ deviate in a non-trivial way cannot be treated so straightforwardly. We shall later see with a practical example what such deviations look like in practice.

\begin{figure}
    \centering
    \includegraphics[width=0.9\linewidth]{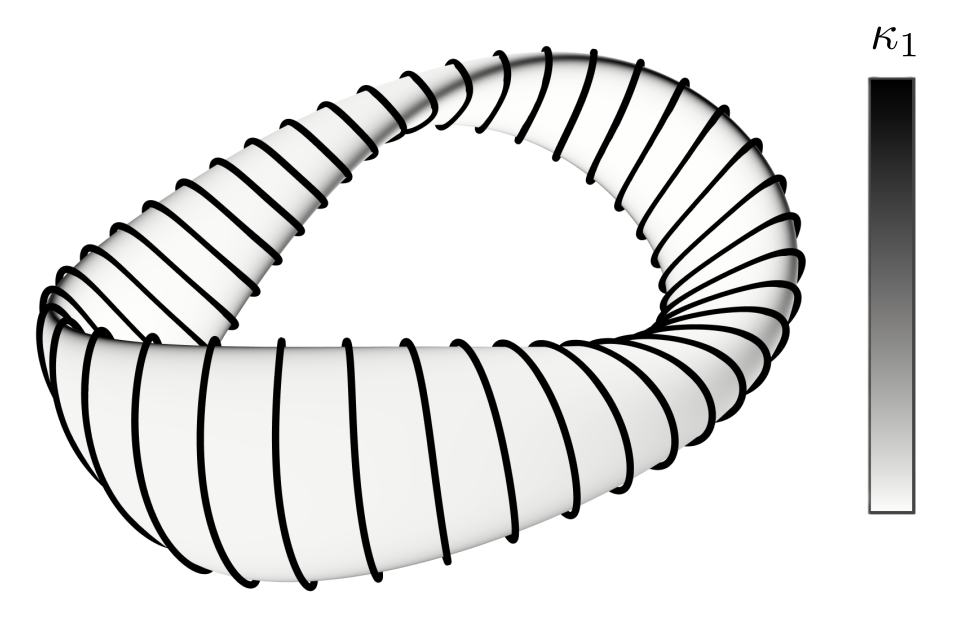}
    \caption{\textbf{Ideal coil set and principal curvature on last close flux surface.} 40-coil construction computed using Eq.~(\ref{eqn:K_anal_B}) for the precise QA stellarator \cite{landreman2021} at $\rho=1$. The surface colour represents the largest principal curvature of the flux surface, indicating the central role of surface curvature on controlling the coil curvature.}
    \label{fig:placeholder}
\end{figure}

In this construction, then, the current potential is precisely the restriction of the vacuum scalar potential to the boundary, $\Phi = \Phi_0|_{\partial\Omega}$ (up to a constant), and is unique. In fact, this is just a scaled version of the Boozer toroidal $\varphi$ angle. If we then interpret this current density, as it is customary, as the limit of a set of filamentary coils (see an example in Figure~\ref{fig:placeholder}), then the coils must be modular; \textit{i.e.}, they close poloidally. { This makes these coils regularised in a sense, given that contours of $\varphi$ always satisfy $\mathbf{B}\cdot\nabla\varphi\neq 0$, preventing them from turning around and other pathological behaviour.}

% A filamentary coil can be thought of as a current carrying closed 1D curve. In that vain, one may think of the surface current $\mathbf{K}$ as the equivalent to an infinite set of 1D coils following contours of constant $\Phi$ and carrying a current equivalent to $\delta\Phi$.

\textit{Coil geometry and intrinsic complexity.}
The geometry of the resulting coils provides direct insight into their intrinsic complexity and how the equilibrium properties influence it. For coils defined as contours of constant $\Phi$ on a flux surface, we can express their curvature as
\begin{gather}
\tilde{\kappa}_n = 2H - \kappa_n,
\qquad
\tilde{\kappa}_g = \frac{1}{B}\,\hat{\tau}\hat{\tau}:\nabla\mathbf{B}, \label{eqn:curvature}
\end{gather}
where $\tilde{\kappa}_g$ and $\tilde{\kappa}_n$ are the geodesic and normal curvatures of the coils, $\kappa_n$ the curvature of the magnetic field lines, $\hat{\pmb \tau}=\mathbf{B}\times\hat{\pmb n}/|\mathbf{B}|$ and $H$ is the mean curvature of $S$. \cite{Struik1988Euler} The total curvature of the coils is then by definition $\tilde{\kappa} = (\tilde{\kappa}_g^2+\tilde{\kappa}_n^2)^{1/2}$.

The normal curvature is primarily set by the surface geometry. Because MHD stability generally pushes magnetic field lines to have as negative a normal curvature as possible, but also due to their tendency to minimise bending energy, coils which are orthogonal to them tend to align with the direction of largest principal curvature $\kappa_1$ (\textit{i.e.}, poloidally). The role of $\kappa_1$ is illustrated in Figure~\ref{fig:placeholder}, and clearly shows why sharp surface features or extreme elongation are to be avoided. Projecting this behaviour outwards, we expect the normal curvature of the coils to drop as roughly $\tilde{\kappa}_n\sim1/\rho$, the normalised radial flux coordinate, as the winding surface is pushed away from the plasma. The particulars of the winding surface shape will of course matter.

\begin{figure}
    \centering
    \includegraphics[width=\linewidth]{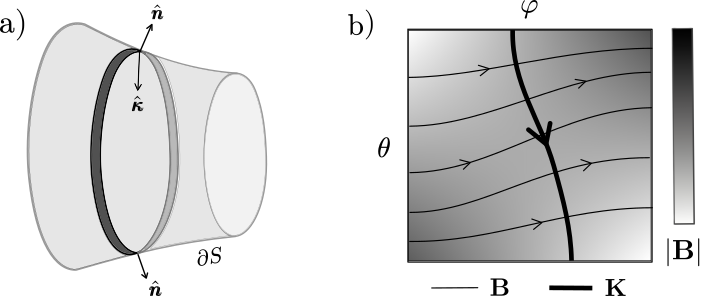}
    \caption{\textbf{Illustrations of geodesic curvature and non-planar behaviour of coils.} a) Diagram showing a flux surface $\partial S$ in the presence of a significant breathing, \textit{i.e.}, toroidal variation due to change in toroidal magnetic field magnitude. It clearly shows the misalignment between curvature of the coil and the surface, and thus a significant geodesic curvature. b) Sketch of a flux surface patch with a varying field strength in (both) $(\theta,\varphi)$. Fieldlines of $\mathbf{B}$ bunch where the field is stronger, leading to the perpendicular current $\mathbf{K}$ having a clear $S$-like excursion. The normal is taken off the page for this illustration.}
    \label{fig:illustrations}
\end{figure}

The geodesic curvature is the other component of curvature, and represents bending of the coil within the surface. This is controlled by a component of $\nabla\mathbf{B}$ describing the bending of field lines in the direction of the coils. Picturing this bending as principally a misalignment of the surface normal to a toroidal cut, significant geodesic curvature naturally arises in configurations with large toroidal variations in the field strength (see Figure~\ref{fig:illustrations}a) or elongation (stretching). Configurations with little toroidal variations minimise this contribution, as is the case of Figure~\ref{fig:placeholder}. 

\begin{figure*}
    \centering
    \includegraphics[width=0.8\linewidth]{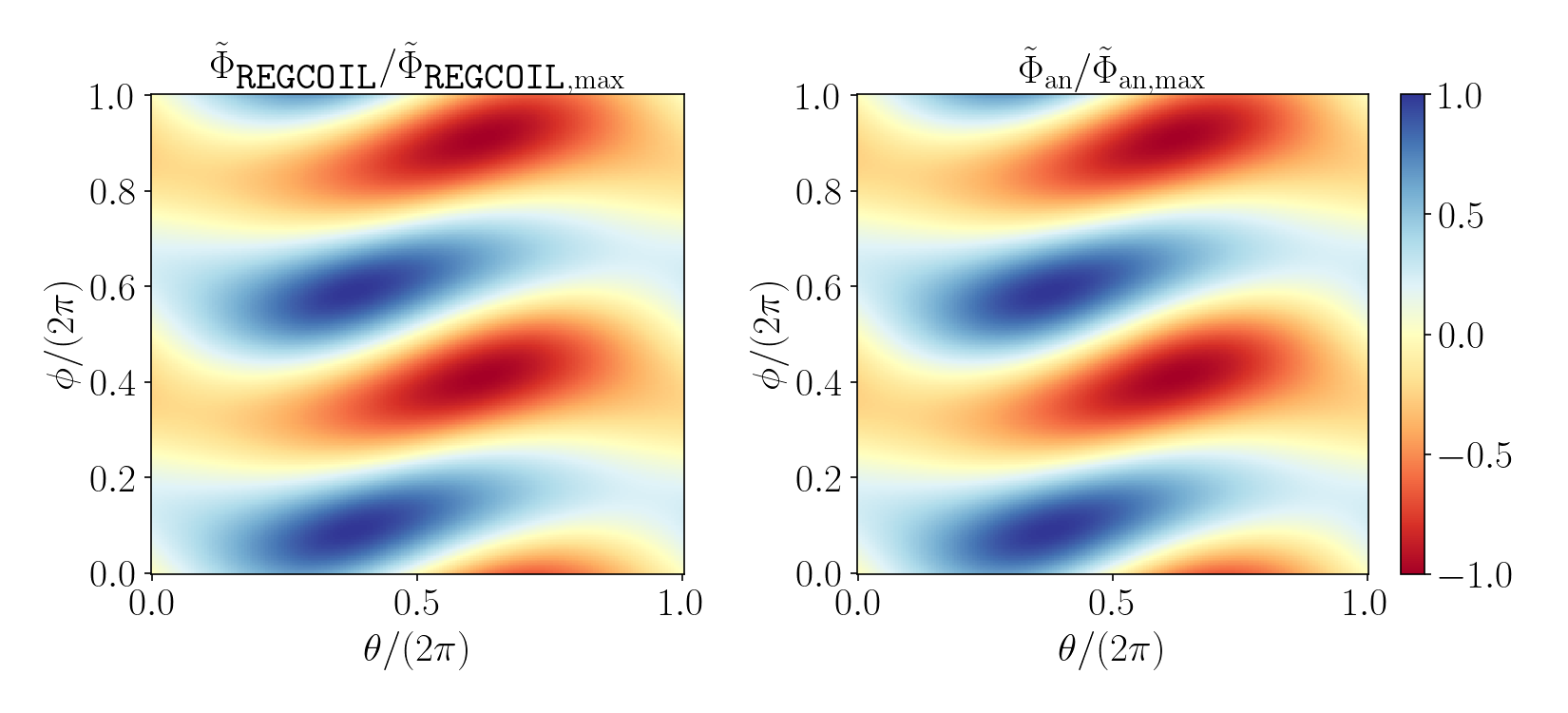}
    \caption{\textbf{Comparison of current potential at $\rho=1$.} The plot shows a comparison of the current potential as computed by the least squares \texttt{REGCOIL} (with regularisation $\lambda=0$) solve and reconstruction using the \texttt{VMEC} field, showing that the asymptotic limit is as theoretically shown, recovered. The numerical \texttt{REGCOIL} construction can only be done approximately, as the singular behaviour of Biot-Savart demands inordinate resolution when the winding and evaluation surface become very close. Here, we have used $\Delta\rho\sim0.03$, with a total of $N_\theta\times N_\varphi=4\times10^6$ collocation points on the winding surface and $M_\Phi\times N_\Phi=20\times20$ Fourier modes for $\Phi$.}
    \label{fig:comparison_Phi}
\end{figure*}

Another important aspect of the coils we have not referred to yet is the non-planarity of coils. The concept could be discussed through the coil torsion, but torsion is a rather sensitive quantity, especially if the curves have straight sections. Thus we opt in this letter for a more direct physical picture leaning on the bending of field lines over flux surfaces. As pictured in Figure~\ref{fig:illustrations}b, a simultaneous poloidal and toroidal variation in the strength of the magnetic field over a flux surface will make field lines bend, and thus following Eq.~(\ref{eqn:K_anal_B}), coils will as well. This picture suggests{, qualitatively,} that stellarator fields that minimise toroidal variations in the poloidal behaviour { (\textit{e.g.}, avoiding large toroidal variations in the cross-section shape normal to the magnetic axis)} will be amenable to { more} planar coils, consistent with recent work on the figure-8 stellarator\cite{plunk2025back} and providing a criterion for designing simpler fields. 

Making this somewhat more quantitative, the toroidal topology, together with Amp\'{e}re's law, forces a strong inboard-outboard variation of the poloidal field, with a basic $\sim\rho$ scaling. Since the coil length scales as $L\sim\rho$, the accumulated out-of-surface excursion is then expected to scale as $\Delta z\sim\rho^2$. Coil non-planarity can thus be expected to roughly grow quadratically with radius. This strong increase can also be understood by invoking the superposition principle to describe equivalently a set of 3D coils with finite $\Delta z$ excursion as a set of planar modular coils together with additional non-linked saddle coils. The saddle coil field being that of dipoles decays rapidly (in fact as $\sim1/\rho^3$, or more appropriate here $1/\rho^2$ if elongated), and thus the excursion it represents must therefore grow strongly. The basic scaling can be exploited to extrapolate the non-planarity of coils away from the plasma on the basis of a direct assessment of our equilibrium field. { We thus can construct a proxy of coil non-planarity by calculating the slope of $\Delta z$ from the equilibrium straightforwardly (and without having to design a finite working set of coils).}

There is therefore a tight competition between the curvature and the non-planarity of the coils as the coil set is placed further and further away from the plasma. A more detailed description is left for future work. 

\textit{Numerical non-planarity assessment.}
To provide some practical support for the scaling arguments above, with particular focus on the non-planarity of coils, we consider an example optimised stellarator, the precise QA equilibrium\cite{landreman2021}, and apply our construction to it. { We also include precise QH\cite{landreman2021} and W7X\cite{wanner2001design} to show the illustrate applicability of the construction.}

First, we apply our explicit construction to the nested set of flux surfaces obtained from \texttt{VMEC} \cite{hirshman1983}. Given the covariant representation of the magnetic field,
\begin{gather}
B_\theta = \partial_\theta \Phi, \qquad
B_\phi = \partial_\phi \Phi,
\end{gather}
we may solve directly for the periodic part $\tilde{\Phi}$ of the potential $\Phi = \tilde{\Phi} + G\phi + I\theta$. { This can be done directly using the spectral representation of the covariant field, integrating both equations and matching the result.} Here, $(G,I)$ are the Boozer currents, with $I=0$ in our vacuum. In Figure~\ref{fig:comparison_Phi}, we compare the current potential reconstruction this way against a numerical solution obtained using \texttt{REGCOIL}, showing the right asymptotic behaviour. 

\begin{figure*}
    \centering
    \includegraphics[width=0.8\linewidth]{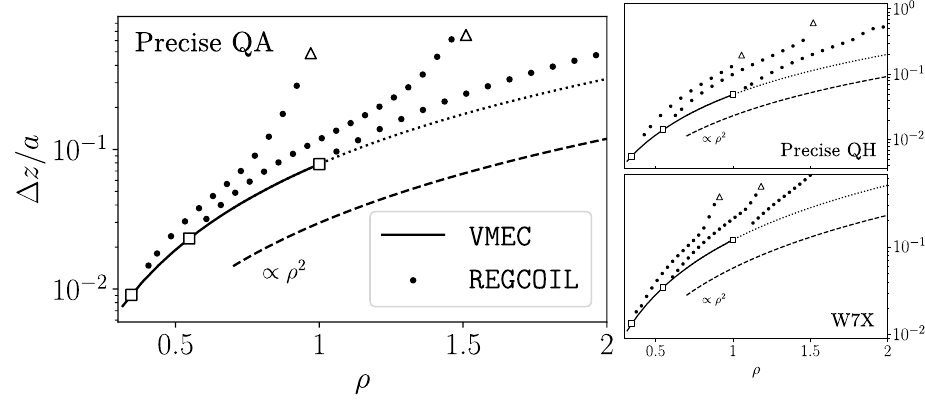}
    \caption{\textbf{Coil non-planarity assessment.} The plot shows a comparison of the coil non-planarity, $\Delta z$ (normalised to the minor radius $a$ at $\rho=1$), as predicted by applying our closed form current density from the \texttt{VMEC} equilibrium (solid black line) and its extrapolation (dotted line), against the solution obtained from \texttt{REGCOIL} (scatter), { for three different stellarators: precise QA\cite{landreman2019}, precise QH\cite{landreman2019} and W7X\cite{wanner2001design}}. Each of the three groups of \texttt{REGCOIL} solves { in each panel} is computed considering the normal deformation of the reference \texttt{VMEC} surfaces at normalised radii $\rho=0.35,0.55,1.0$ (squares) by different amounts $d$, up to when modular coils are no longer a solution (triangles). The effective $\rho$ is then defined as $1+d/a$, where $a$ is the minor radius for the \texttt{VMEC} equilibrium at $\rho=1$.}
    \label{fig:regcoil_extrapolate}
\end{figure*}

We can then compute the non-planarity $\Delta z$ of the constructed coil sets by defining, for each constant $\Phi$ contour, a least-squares plane and computing the largest normal deviation from it. Our construction can be evaluated on every flux surface within $\rho\leq1$, and is shown in Figure~\ref{fig:regcoil_extrapolate} as a solid line. The dotted line is its $\rho^2$ extrapolation. To compare it to \texttt{REGCOIL}, we choose three reference surfaces at different $\rho$, and solve the Merkel problem at a set of surfaces constructed by a uniform outward-displaced of each reference surface. Within $\rho\leq1$, it is clear that the standard \texttt{REGCOIL} choice of winding surface leads to a coil excursion bounded from below by our construction, which in these cases minimise the excursion. This difference is mainly due to the winding surface shaping, which leads to a breakdown of the modular coil assumption for the \texttt{REGCOIL} construction beyond a critical $\rho$ (see triangular scatter in Figure~\ref{fig:regcoil_extrapolate}). This effect lies outside the scope of the present framework and is clearly not intrinsic to the equilibrium.

Beyond the last closed flux surface, the quadratic extrapolation appears to continue to provide a lower bound on coil non-planarity. A more thorough exploration remains to be carried out, but we hypothesise that a suitably deformed surface, { appropriately regularised current distribution}, or more freely moving filamentary coils, will be able to approach this limit. Techniques for extending the flux-surface solution outward\cite{liu2026near} and provide pseudo-flux-surfaces (or flux minimising surfaces) may thus be of practical interest { to provide initial coil sets for further optimisation}. In any event, we propose the use of this equilibrium based excursion measure as a guiding design feature for simplifying coils. 

\textit{Conclusions.}
In this Letter, we have introduced a construction in which a coil set is defined uniquely for a given stellarator equilibrium by placing the current potential on a flux surface. This removes the intrinsic non-uniqueness of the coil design problem and yields a direct correspondence between magnetic-field structure and coil geometry. Within this framework, key geometric properties of the coils, such as curvature and non-planarity, can be related explicitly to local properties of the magnetic field.

The theoretical construction can serve as a lower-bound estimate for the excursion of practical coil designs. First comparisons with more realistic approaches indicate promising predictive power, which merits further study. More broadly, the present approach opens the possibility of systematically identifying equilibria that admit simpler coil sets, understanding them more deeply, and formulating quantitative criteria for coil simplicity directly in terms of magnetic-field structure.

\par

\section*{Acknowledgements}
The authors would like to acknowledge fruitful discussion with Michael Drevlak, Annika Zettl, Per Helander, Andrea Pavone and Gabriel Plunk, Elizabeth Paul, Mike Zarnstorff, John Kappel, Sanket Patil, Pedro Gil, Lanke Fu, and Matt Landreman.

\section*{Author declaration}
The authors have no conflicts to disclose.

\section*{Data availability}
The data that support the findings of this study are openly available in GitHub, with a Zenodo DOI https://doi.org/10.5281/zenodo.21197246.

\appendix

\bibliography{condQS}% Produces the bibliography via BibTeX.

\end{document}